\documentclass[12pt]{iopart}
\usepackage{epsf}
\def\simlt{\lower.5ex\hbox{$\; \buildrel < \over \sim \;$}}

\begin{document}

\title[Extracting SZ parameters]
{Extraction of cluster parameters with future Sunyaev-Zel'dovich
observations}

\author{Nabila Aghanim\dag, Steen H.\ Hansen\ddag, 
Sergio Pastor\S\ and Dmitry V.\ Semikoz$\|$\P}

\address{\dag\ IAS-CNRS, B\^atiment 121, Universit\'e Paris Sud
F-91405 Orsay, France}

\address{\ddag\ Institute for Theoretical Physics, Univ. of Zurich,
Winterthurerstrasse 190, CH-8057 Zurich,  Switzerland}

\address{\S\ Institut de F\'{\i}sica Corpuscular (CSIC/Universitat
de Val\`encia), Ed.\ Institutos de Investigaci\'on, Apdo.\ 22085,
46071 Valencia, Spain}

\address{$\|$ Max-Planck-Institut f\"ur Physik
(Werner-Heisenberg-Institut), F\"ohringer Ring 6, 80805 Munich,
Germany}

\address{\P\ Institute of Nuclear Research of the Russian Academy of Sciences,
60th October Anniversary Prospect 7a, Moscow 117312, Russia}

\begin{abstract}
The Sunyaev-Zel'dovich (SZ) effect of galaxy clusters is characterized
by three parameters: Compton parameter, electron temperature and
cluster peculiar velocity. In the present study we consider the
problem of extracting these parameters using multi-frequency SZ
observations only. We show that there exists a parameter degeneracy
which can be broken with an appropriate choice of frequencies. As a
result we discuss the optimal choice of observing frequencies from a
theoretical point of view. Finally, we analyze the systematic errors
(of the order $\mu$K) on the SZ measurement introduced by finite
bandwidths, and suggest a possible method of reducing these errors.
\end{abstract}



\eads{\mailto{Nabila.Aghanim@ias.u-psud.fr},
\mailto{hansen@pegasus.physik.unizh.ch}, \mailto{pastor@ific.uv.es},
\mailto{semikoz@mppmu.mpg.de}}

\maketitle

\section{Introduction}

The interaction of the Cosmic Microwave Background Radiation (CMBR)
photons with the free electrons in the ionized gas of clusters of
galaxies produces a small change in the intensity known as the
Sunyaev-Zel'dovich (SZ) effect. This distortion arises from the
transfer of photons from the low-energy to the high-energy or Wien side of
the planckian spectrum. The underlying physics of the SZ effect is
well understood and a quantitative description was given by Sunyaev \&
Zel'dovich \cite{sz}, who already realized its cosmological
significance. Nowadays there exists reliable observations of the SZ
effect (see e.g.~\cite{LaRoque01,coma,joy2001})
and dedicated experiments are being prepared, which will
provide information about the kinematics and evolution of clusters,
which in turn can be used to extract cosmological parameters (see
e.g.~\cite{rep95,bir99,car02} for recent reviews).

The intensity change of the CMBR caused by the SZ effect is
proportional to the Comptonization parameter,
\begin{equation}
y_c = \int dl~ \frac{T_e}{m_e} ~ n_e \sigma_{\rm Th} ~,
\end{equation}
where $T_e$ is the temperature of the electron gas in the
cluster, $m_e$ the electron mass, $n_e$ the electron number density,
$\sigma_{\rm Th}$ the Thomson scattering cross section, and the
integral is calculated along the line of sight through the cluster. We
use units for which $k_B=\hbar=c=1$.  For an isothermal intra-cluster
gas one has $y_c=\tau \, T_e/m_e$, where $\tau=\int dl~ n_e
\sigma_{\rm Th}$ is the optical depth. The distortion of the CMBR is
then given by an intensity change
\begin{equation}
\Delta I_{\rm T} = 
I_0~ y_c~ \left( f_0(x)~ +\delta f(x,T_e)\right)~,
\label{thermal}
\end{equation}
where
\begin{equation}
f_0(x) = \frac{x^4e^x}{(e^x-1)^2}\left[
\frac{x(e^x+1)}{e^x-1}-4\right]~.
\label{f0}
\end{equation}

\begin{figure}
\begin{center}
\epsfxsize=9cm
\epsffile{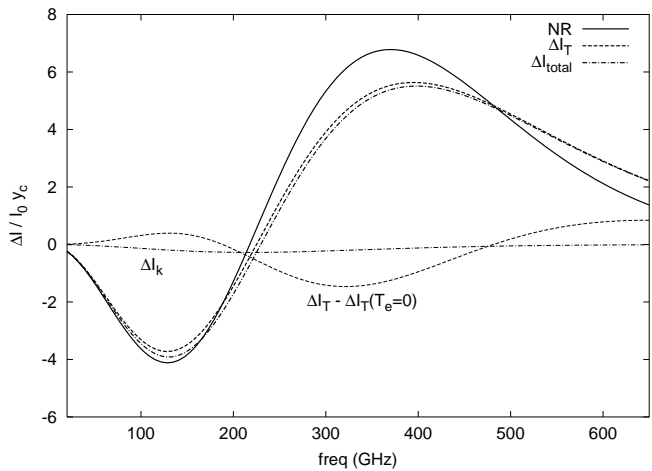}
\end{center}
\caption{The normalized intensity change caused by the thermal and
  kinetic SZ effects for the choice $T_e=15$ keV and $v_p=500$ km/sec.
  The solid line is the non-relativistic SZ effect. The dashed lines
  are the relativistic corrections, both alone and together with the thermal
  effect.  The dot-dashed lines are the kinetic SZ effect, both alone
  and together with the thermal effect.}
\label{fig:example}
\end{figure}

Here $x=\nu/T_0$ is the dimensionless frequency with $T_0 =2.725$ K,
and $I_0=T_0^3/(2\pi^2)$. The intensity change is independent of the
temperature for non-relativistic electrons, a limit which is valid for
small frequencies ($\nu\simlt 100$ GHz), but for high frequencies it
must be corrected \cite{rep95b} with $\delta f(x,T_e)$ either by using
an expansion in $\theta_e=T_e/m_e$ \cite{cha98,ito98,noz00} or
calculated exactly \cite{dol01}. In some rich clusters the optical
depth is large enough ($\tau \sim 0.02-0.03$) to require the inclusion
of multiple scattering effects \cite{dol01,ito01,colaf}.

When the cluster has a relative motion with respect to the CMBR rest
frame, there exists an additional distortion known as the kinetic SZ
effect \cite{sz80}, which is identical to a change in the temperature
of the CMBR.  The corresponding intensity change, for a peculiar
velocity $v_p$, is $\Delta I_{\rm K} = I_0 y_c f_{\rm kin}$, where 
\begin{equation}
f_{\rm kin} = -   v_p \frac{m_e}{T_e}
 \frac{x^4 e^x}{(e^x-1)^2}~. 
\label{kinematic}
\end{equation}
This expression is slightly modified by the corrections caused by the
electron temperature \cite{saz98,noz98}.

The total intensity change is just the sum of the
thermal and kinetic SZ effects,
\begin{equation}
\Delta I_{\rm total} = \Delta I_{\rm T}+\Delta I_{\rm K}~,
\label{total}
\end{equation}
and it is characterized by three parameters: $y_c, T_e$ and $v_p$.
The different contributions to the SZ distortion are presented in
\fref{fig:example}, where the total intensity change in \eref{total}
is compared to the thermal effect with vanishing temperature (the
first term in \eref{thermal}). The kinetic SZ effect and the
contribution of the relativistic corrections (the second term in
\eref{thermal}) are presented for the parameter choice $T_e=15$
keV and $v_p = 500 ~{\rm km}/{\rm s}$.

Let us emphasise that all 3 SZ parameters can in principle be
extracted from multi-frequency observations of the SZ effect. This is
because the frequency dependence of the SZ effect is different for the
different parameters. This is clear from \fref{fig:example}, where one
can see that the main effect (the non-relativistic SZ effect
characterized by the comptonization parameter, $y_c$) has one crossing
with the zero-line. The kinetic effect (governed by $v_p$) has
no crossings of the zero-line, and the relativistic effect (governed
by $T_e$) crosses the zero-line twice. Thus with several observing
frequencies placed optimally and with good sensitivity, one can
extract all 3 SZ parameters as we describe in detail in \sref{optimal}.

In this paper we discuss the possibilities for future SZ observations
to extract these 3 cluster parameters, taking into account the
parameter degeneracies described in \sref{sec:deg}. To this
end we try to extract the 3 SZ parameters from a set of simulated
galaxy clusters with realistic temperatures and peculiar velocities,
using the sensitivities of upcoming experiments like ACT and
Planck. In \sref{optimal}, we discuss the optimal observing
frequencies from a theoretical point of view. Finally, we analyze in
\sref{bandwidth} the systematic error introduced by a finite
frequency bandwidth, which may be as large as few $\mu$K and hence
important for the next generation of SZ observations.

\section{Parameter degeneracies}
\label{sec:deg}
When one estimates the expected error-bars of a given experiment,
it is always important to consider the effect of parameter
degeneracies.  This is naturally also true when considering the SZ
effect. Let us clarify that by degeneracies we mean that two sets of
parameters give an indistinguishable intensity change due
to the limited sensitivity of the observations. In other words, the
extracted values of $y_c, v_p$ and $T_e$ from the same experiment will
have larger error bars for some of the clusters than for others.

\begin{figure}
\begin{center}
\epsfxsize=9cm
\epsffile{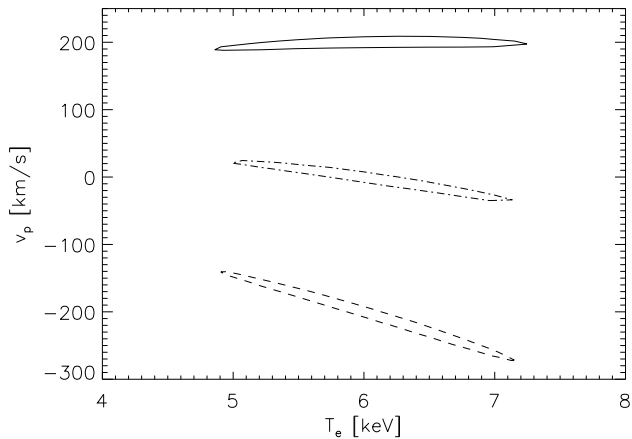}
\end{center}
\caption{$2\sigma$ contour plots for an experiment
  with 4 frequencies at 90, 181, 220 and 330 GHz, with 1 $\mu$K
  sensitivity.  The 3 clusters have the same temperature $T=6$ keV and
  comptonization parameter, $y_c=2\times 10^{-4}$, but have different
  peculiar velocities, $v_p = -200, 0,+200$ km/sec.}
\label{fig:3contours}
\end{figure}

The degeneracy between the 3 SZ parameters depends on the chosen value
of the parameters, e.g. if the peculiar velocity is large and
positive, then it is virtually impossible to adjust the temperature
and comptonization parameter to mimic the effect on the intensity
change of the peculiar velocity.  On the other hand, for large and
negative peculiar velocities, such imitation is possible.
\begin{figure}
\begin{center}
\epsfxsize=9cm
\epsffile{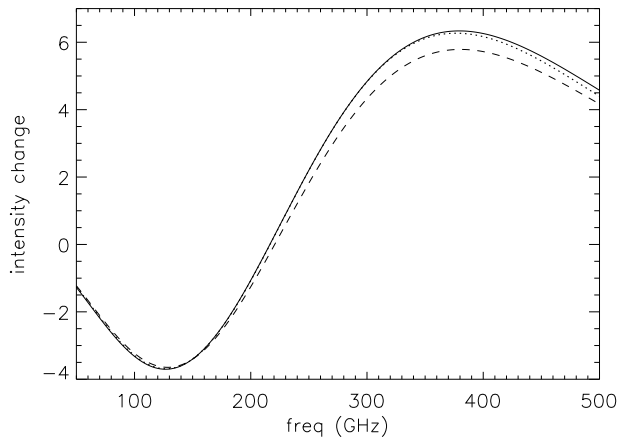}
\end{center}
\caption{The intensity change for 3 different clusters, normalized at 
  150 GHz. The solid line is $T=5$ keV, $v_p=-145$ km/sec and the dotted
  is $T=7$ keV, $v_p=-260$ km/sec. These two are virtually
  indistinguishable. The dashed line is $T=6$ keV, $v_p=0$ km/sec, which
  differs significantly for large frequencies.}
\label{fig:deg}
\end{figure}
In \fref{fig:deg} the solid and dotted lines are virtually
indistinguishable.  They are with $T=5$ keV, $v_p=-145$ km/sec and
$T=7$ keV, $v_p=-260$ km/sec respectively, the comptonization
parameter is normalized to make the intensity change agree at 150 GHz.
For comparison we see, that the dashed line, $T=6$ keV, $v_p=0$
km/sec, also normalized to agree at 150 GHz, can more easily be
distinguished.  As a specific example, let us consider a hypothetical
experiment, with 4 frequencies at 90, 181, 220 and 330 GHz, with 1
$\mu$K sensitivity.  With 3 test clusters all with the same
temperature, $T=6$ keV and comptonization parameter, $y_c=2\times
10^{-4}$, but with different peculiar velocities, $v_p = -200, 0,+200$
km/sec. The resulting $2\sigma$ error-contours are given
\fref{fig:3contours}, and it is clear from the figure, that
whereas the temperature error-bars are virtually independent of the
real value of $v_p$ (always about $\pm 1$ keV), then the error-bar of
$v_p$ is strongly dependent, and may differ from $10$ km/sec for
$v_p=+200$ km/sec to $70$ km/sec for $v_p=-200$ km/sec, i.e. by a
factor of 7.  Hence it is important to consider a range of
cluster parameters when one wants to make predictions about future
experiments.

In this paper we therefore use realistic simulations of clusters
of galaxies (described below), and the error-bars will not be for
a specific cluster, but instead presented for a large samples 
of clusters.

\section{Simulated clusters}
\label{sec:simu}

In order to produce a large sample of clusters with realistic temperatures,
comptonization parameters and peculiar velocities, one must basically use
an extended Press--Schechter formalism. Let us here discuss the details
of the used simulation which closely follows~\cite{agh01}.

The physical parameters (temperature, core and virial radii, central
electron density, ...) used to describe the clusters are computed
according to their mass and redshift. In the following, we use the
most favoured cosmological model with critical density described by
the following 
cosmological parameters: the reduced Hubble constant $h=0.70$, 
cold dark matter density parameter
$\Omega_{cdm}=0.12h^{-2}$, baryonic density parameter
$\Omega_{b}=0.02h^{-2}$, total matter density parameter
$\Omega_M=\Omega_{cdm}+\Omega_{b}$, and a 
cosmological constant $\Omega_{\Lambda}=1-\Omega_M=
0.714$.
\par The cluster core radius $R_c$ is related to the virial radius
$R_v$ through the parameter $p=R_{vir}/R_c$, where we use 
$p=10$~\cite{arnaud2001}.  The
virial radius is given by the following equation
\begin{equation}
R_{vir}=\frac{(G\,M)^{1/3}}{(3\pi\,H(z))^{2/3}},
\end{equation}
where $G$ is the gravitational constant and $H(z)$ the Hubble constant.
For a critical universe, $R_v$ is fixed solely by the mass and the collapse 
redshift of the cluster,
\begin{equation}
R_{vir}=\frac{(G\,M)^{1/3}}{(3\pi\,H_0)^{2/3}}\,\frac{1}{1+z},
\end{equation}
with $H_0=100\times h$ the value of the Hubble constant today.
The electron density profile for the clusters is well described by the
so-called phenomenological $\beta$ profile as follows
\begin{equation}
n_e(R)=n_{e0}\left[1+\left(\frac{R}{R_c}\right)^2\right]^{-3\beta/2},
\label{eq:prof}
\end{equation}
where $n_0$ is the central electron density and $\beta$ is a parameter 
describing the steepness of the profile. The X-ray brightness profiles of 
galaxy clusters mostly agree with $\beta=2/3$, which is the value we use.
Finally knowing the virial radius, we can derive the central
electron density $n_0$ from the gas mass of the cluster $M_G$ using the
following relation
\begin{equation}
M_G\left(\frac{\Omega_b}{\Omega_M}\right)=m_p\mu\int_0^{R_v}n_e(R)
\,4\pi R^2\,dR. 
\end{equation}
In this relation, $m_p$ is the proton mass and $\mu=0.6$ is the mean
molecular weight of a plasma with primordial abundances. In our model,
we use a gas fraction of $0.166(h/0.7)^{-1}$
(where the gas fraction
equals the baryon fraction, and is
slightly higher than some observed values~\cite{grego2001}).
We compute the
temperature of the intra-cluster medium as a function of the mass, the
redshift and the cosmological model
\begin{equation}
T=\frac{GM^{2/3}\mu m_p}{2\beta}\left[\frac{H^2(z)\Delta_c}{2G}\right]^{1/3},
\end{equation}
and thus
\begin{equation}
T=1.39f_TM_{15}^{2/3}(h^2\Delta_cE^2(z))^{1/3},
\end{equation}
where $T$ is given in keV and $M_{15}$ is the mass of the structure
in units of $10^{15}$ solar masses. $f_T=0.78$ is a normalization
factor, we take it as a unique value from Bryan \& Norman
\cite{bryan98} although it varies very slightly with the
cosmological model. $\Delta_c$ is the average over-density
($\Delta_c=18\pi^2$ for a critical universe) and $E^2(z)$ is a
function such as $H^2(z)=H_0^2\,E^2(z)$ which depends on the
cosmology. \par The number density of clusters are derived from a
modified Press--Schechter mass function according to Sheth \& Tormen
\cite{sheth99} and Wu \cite{wu01}. The numbers are
normalized to Viana \& Liddle \cite{via99} and we use
$\sigma_8=\sigma_8^0\Omega_m^{-0.47}$ with $\sigma_8^0\,=\,0.6$. \par
On the scale of clusters of galaxies we assume that the density
fluctuations are in the linear regime.  In fact, under the assumption of
an isotropic Gaussian distribution of the initial density
perturbations, the initial power spectrum $P(k)$ gives a complete
description of the velocity field through the three--dimensional {\it
rms} velocity ($v_{rms}$) predicted by the linear gravitational
instability for an irrotational field at a given scale $R$
\cite{pee93}. This velocity is given by
\begin{equation}
v_{rms}=a(t)\,H\,f(\Omega,\Lambda)\left[\frac{1}{2\pi^2}\int^{\infty}_0 P(k)
W^2(kR)\,dk\right]^{1/2}
\label{vrms:eq}
\end{equation}
where $a(t)$ is the expansion parameter, the Hubble constant $H$ and
the density parameter $\Omega$ vary with time \cite{car92}. The
function $f(\Omega,\Lambda)$ is accurately approximated by
$f(\Omega,\Lambda)=\Omega^{0.6}$ \cite{pee80}.  Furthermore, under the
assumptions of linear regime and Gaussian distribution of the density
fluctuations, the structures move with respect to the global Hubble
flow with peculiar velocities following a Gaussian distribution
$f(v)=\frac{1}{v_{rms}\sqrt{2\pi}}\exp(\frac{-v^2}{2v_{rms}^2})$ which
is fully described by $v_{rms}$. This prediction is in agreement with
numerical simulations. The velocity of each cluster (at mass $M$
associated with the scale $R$) is then randomly drawn from the
Gaussian distribution.

Thus we have produced a large sample of realistic galaxy clusters
each with their {\em true} comptonization parameter, temperature and
peculiar velocity, and we will in the next sections 
consider this sample as being what can be observed in future all-sky 
observations.

\section{General method}
In this section, we present the method used to answer the following
question: How well can a multi-frequency experiment measure the
physical cluster parameters (namely $T_e$, $v_p$ and $y$)? For a given
simulated cluster, we have the {\em true} value of the parameters,
$T_e^{\bf t},v_p^{\bf t}$ and $y_c^{\bf t}$.  With these values we can
calculate what the {\em true} SZ signal should be at each frequency,
and then the sensitivity of the experiment gives us an estimate of the
expected observational error-bars at each frequency.  Now given
this {\em observed} SZ signal (with its error-bars), we can try to {\em
deduce} the 3 parameters $T_e^{\bf d},v_p^{\bf d}$ and $y_c^{\bf d}$,
and corresponding error-bars,
and we can then compare with the true values, $T_e^{\bf t},v_p^{\bf t}$ and
$y_c^{\bf t}$.

To deduce the parameters we have developed a method which in 3 steps
finds the central value and 1$\sigma$ error-bars of each
parameter. The code is an extension of the method developed in Hansen
et al.\ \cite{han02}.  First, we find the approximate $5\sigma$
region in $y_c$ (in the range $10^{-6.5} < y_c < 10^{-1.5}$). Next, we
define a grid for all 3 parameters (using the above mentioned
$y_c$-range, and $0<T<20$ keV and $-2000<v_p< +2000$ km/sec), and we
find an approximate 3$\sigma$ range for all parameters. Finally, the
code makes a new refined grid in the ranges derived from the previous
step (3$\sigma$ range), and calculates the central values and
1$\sigma$ error-bars for $T_e$, $v_p$ and $y_c$. The deduced central
parameters will be unrealistically close to the true values, since we
assume that the true intensity variation due to SZ effect is the
central observational intensity variation. However, the estimated
error-bars will nevertheless be realistic.

In the approach described above, we have assumed that no external
information is known about the observed value. In that case, all three
parameters are derived from the SZ observations. We can relax this
assumption by allowing the temperature $T_e$ to be known at $\pm10\%$
accuracy, e.g. from X-rays observations.
The temperature is in that case in the limited range of $\pm10\%$ around
the {\em real} temperature.  This would be true for a limited sample of
clusters, however, the X-ray observations are time consuming, so one
might not want to rely on them for large cluster surveys.  By doing
this, we estimate the importance of priors on the temperature on the
determination of the deduced parameters.
\section{Application to CMBR and SZ experiments}

\subsection{The ACT experiment}

Let us consider the Atacama Cosmology Telescope (ACT), a proposed
future SZ experiment~\cite{act}.  This experiment will observe at 3
frequencies (150, 220 and 270 GHz), with $0.9-1.7'$ resolution, and an
expected sensitivity of 2 $\mu$K per pixel~\cite{kosowsky}.  The ACT
team expects conservatively to observe $\sim 10^3$ galaxy clusters
through the SZ effect.

In order to test the capabilities of ACT to the determination of the
cluster parameters, we analyze a subsample with the 500 hottest
clusters of our simulation (which with the chosen cosmological
parameters lie in the range $3.6<T_e<7.4$ keV). For each simulated
cluster we have both the {\em true} values, and the ones {\em deduced}
with our method.  It turns out, that ACT will basically not be able to
extract the temperature (within the reasonable range $0<T_e<20$ keV),
partly because of the range of temperatures of our cluster subsample.
One should keep in mind, that e.g. during cluster 
mergers the temperatures will rise significantly, in which case the
relativistic corrections become more prominent, and a temperature
detection would be easier.
Thus we have always left $T_e$ as a free parameter, which must be
marginalized over. We therefore focus on the two other parameters
$y_c$ and $v_p$, for which we have performed two runs, one where the
temperature is known within $10\%$, and one with no prior knowledge of
the temperature.

We find that the comptonization parameter can be deduced with
1$\sigma$ error-bars of $10\%$ when the temperature is unknown,
whereas the results are about $2\%$ with prior temperature
information. These accuracies are obtained for a cluster with
$y_c\approx 10^{-4}$. For a Compton parameter about ten times
smaller the accuracy is still very good $\simeq 15\%$ without prior
and $\simeq 6\%$ if the temperature is known to $10\%$. 
We also notice that the error-bars are not symmetric for the unknown
temperature case, being larger for positive $\delta y_c/y_c^{\bf t}$,
which is because a larger $y_c$ can be compensated by a simultaneous
larger temperature and larger negative peculiar velocity.

The results for the peculiar velocity show a strong effect of the
degeneracies discussed in \sref{sec:deg}, namely positive
peculiar velocities will be extracted with much smaller error-bars
than negative peculiar velocities. In the case with known temperature
(always within $10\%$) this degeneracy is almost completely broken,
the error-bars are almost symmetric, and of the order $10-20\%$ for
most peculiar velocities.

\subsection{An experiment with 4 frequencies}

As we saw above, a SZ experiment measuring at 3 frequencies like ACT
produce large error-bars when trying to extract all 3 SZ parameters
simultaneously. One way around this problem is to have prior knowledge
of the temperature, which thus effectively reduces the number of
parameters to two. This improves the accuracy with which the parameters
are determined. \par
\begin{figure}
\begin{center}
\epsfxsize=9cm
\epsffile{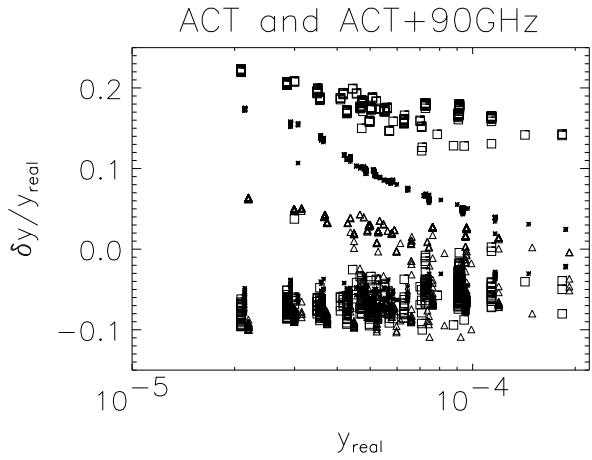}
\end{center}
\caption{The {\em deduced} versus the {\em real} comptonization
  parameter for the upcoming SZ survey ACT with 3 observing
  frequencies 150, 220 and 270 GHz, and an extension of it with one
  extra channel at 90 GHz. The open squares (stars) are the $1\sigma$
  error-bars for ACT (ACT+90) when we have no prior information on the
  SZ parameters.  For comparison, the triangles are the 1$\sigma$
  error-bars when the electron temperature is restricted to the range
  $T_e<5$ keV. }
\label{fig:ACT90y}
\end{figure}

\begin{figure}
\begin{center}
\epsfxsize=9cm
\epsffile{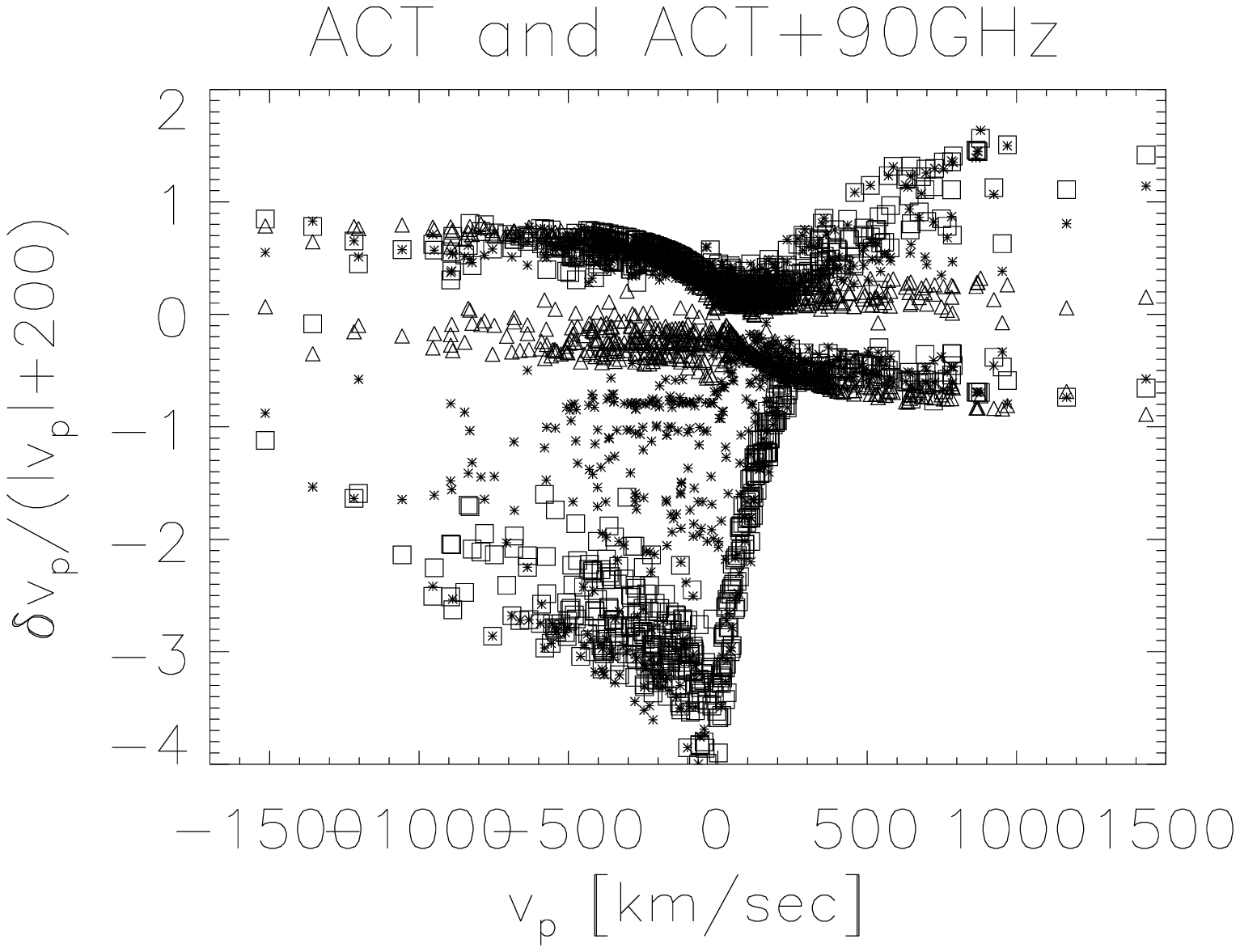}
\end{center}
\caption{Same as \fref{fig:ACT90y} for the peculiar velocity $v_p$.} 
\label{fig:ACT90vp}
\end{figure}

 Another way around this problem is to increase
the number of frequencies of observation. As an example, we show in
figures \ref{fig:ACT90y} and \ref{fig:ACT90vp} the $1\sigma$
error-bars resulting from 3 different runs. Two of these have no prior
information about the SZ parameters.  The first (open squares) is the
same as discussed above, namely 3 frequencies centred on 150, 220 and
270 GHz and sensitivity 2 $\mu$K. The second run (stars) is an
extension of ACT, with one extra observing frequency added at 90
GHz. As can be seen on the figures, the error-bars for both $y_c$ and
$v_p$ become smaller and more symmetric, an indication of breaking of
the degeneracy. Unfortunately, the level of accuracy reached in this
way is far from what was achieved using the prior on the temperature.
This is because contrary to the previous case, one then has to solve
for the values of all three parameters.  One should keep in mind that
it may not be realistic to expect X-ray measured temperatures for that
many clusters, and that additional observing frequencies thus might be
an interesting alternative.  For comparison, we have also calculated
the 1$\sigma$ error-bars when the electron temperature is restricted
to the range $T_e<5$ keV (which is true for most clusters in our
subsample), presented in figures \ref{fig:ACT90y} and \ref{fig:ACT90vp}
with triangles. Making such model-dependent assumption
(temperatures restricted to the range $T_e<5$ keV) is essentially the
same as having prior temperature knowledge.

An alternative to add a new frequency band to the ACT experiment could
be to measure the SZ effect of the same clusters with a different
experiment. An example is the Atacama Large Millimetre Array (ALMA)
project, which has a frequency centred on 35 GHz with precision up to
a few $\mu$K \cite{alma}. Such low frequencies are very good in 
restricting the dominating comptonization parameter.
The main feature of ALMA will be a very good
angular resolution up to scales of a few to tens of arc-seconds, which
will lead to high resolution SZ images.

\begin{figure}
\begin{center}
\epsfxsize=9cm
\epsffile{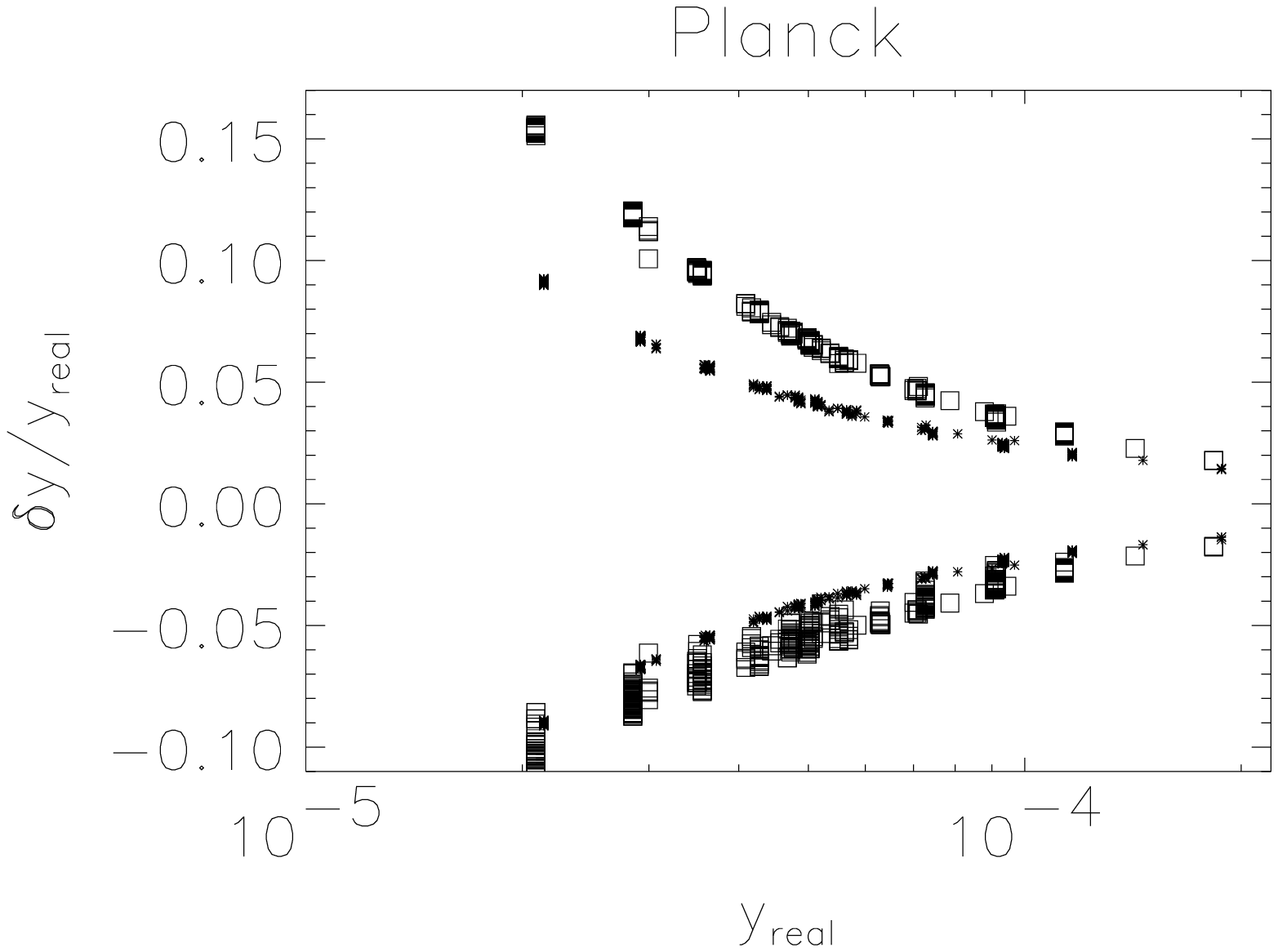}
\end{center}
\caption{The $1\sigma$ error-bar on the {\em deduced} Compton
  parameter $y_c$ for the Planck satellite. The open squares are with
  no prior temperature knowledge, and the stars are when the
  temperature is known within $10\%$.}
\label{fig:plancky}
\end{figure}

\begin{figure}
\begin{center}
\epsfxsize=9cm
\epsffile{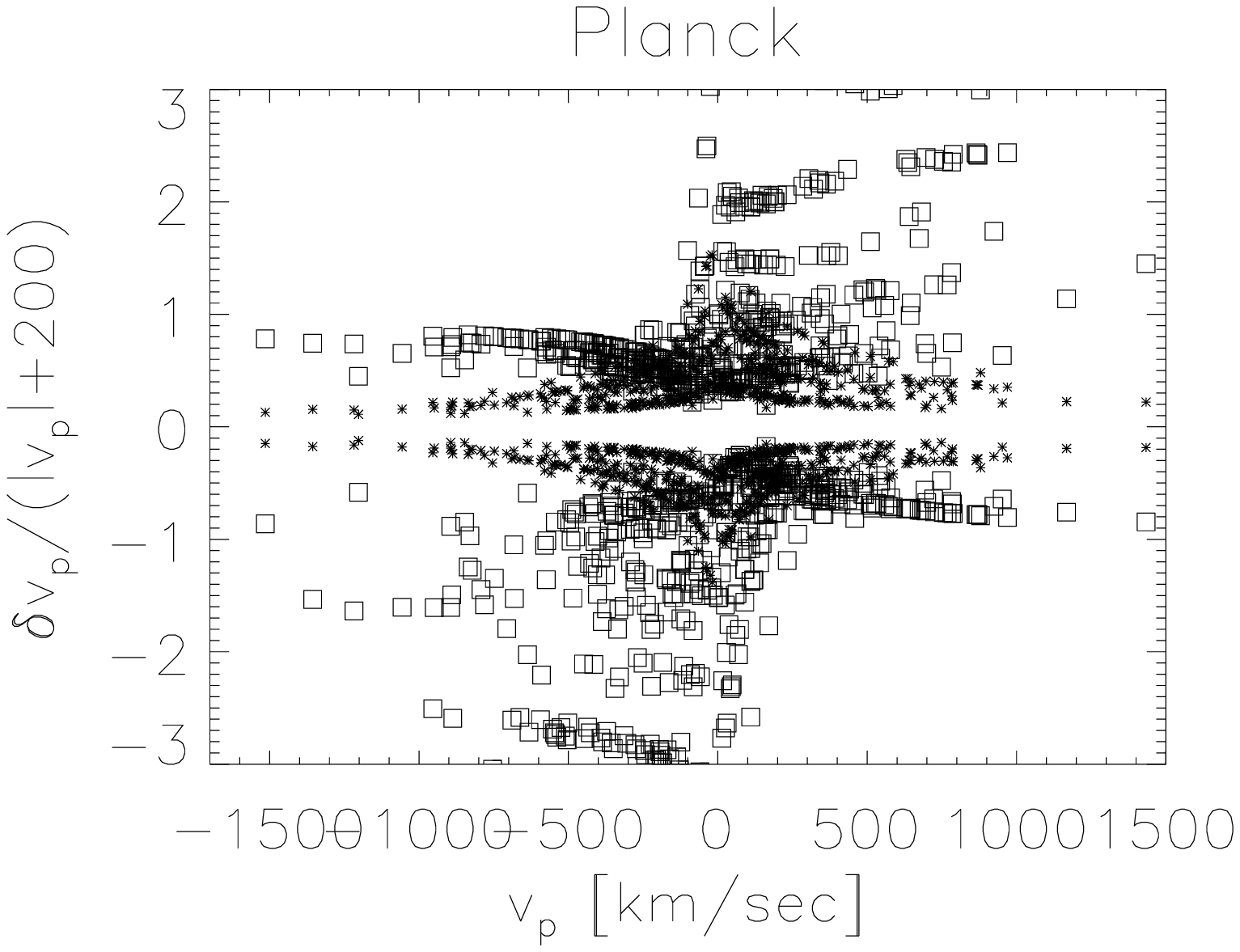}
\end{center}
\caption{Same as \fref{fig:plancky} for the peculiar velocity $v_p$.} 
\label{fig:planckvp}
\end{figure}

\subsection{Planck mission}

The Planck satellite~\cite{planck} will be observing the millimetre
and sub-millimetre wavelengths in nine frequencies. It is thus natural
to investigate how well the three SZ parameters can be extracted with
the Planck sensitivity. Such analysis was first made by Pointecouteau
et al.\ \cite{pgb98}, with the simplifying assumption of
vanishing peculiar velocity.  As discussed above, due to the parameter
degeneracy such simplification should not be made. In the present
study, we rather take explicitly into account the kinetic SZ effect in
order to {\em deduce} all three cluster parameters. It is worth
making the remark that the component separation algorithms for Planck are
so efficient that even though the exact shape of the SZ effect is
unknown (due to relativistic corrections) the SZ contamination
will not be a problem for the CMBR analysis~\cite{diego}.

We assume that the central Planck frequencies (44, 70, 100, 143, 217
and 353 GHz) are observing with sensitivities (2.4, 3.6, 1.7, 2.0, 4.3
and 14.4 $\mu$K). We do not include the lowest and highest frequencies
(30, 545, and 857 GHz) which are dominated by other astrophysical
sources (point sources, dust), and will thus be used to remove the
contaminants. Assuming that the 6 central frequencies have a clean SZ
signal is naturally too ambitious, and our results will therefore
provide an upper bound on the abilities of Planck. Also Planck will
for most clusters not provide any temperature determination, although
we note again that this is partly because of the low temperatures of
our subsample. The $1\sigma$ error-bars on the deduced Compton
parameter and peculiar velocities are shown in figures
\ref{fig:plancky} and \ref{fig:planckvp}.

The comptonization parameter will be determined to about 2\% and
10-15\% for $y_c=2\times10^{-4}$ and $y_c=2\times10^{-5}$
respectively. A 10\% temperature knowledge will only improve these
numbers slightly. The peculiar velocity will only be determined within
a factor of a few, however, a 10\% temperature knowledge will improve
this to about 20-50\%.

\subsection{Extracting the cluster temperatures}

\begin{figure}
\begin{center}
\epsfxsize=9cm
\epsffile{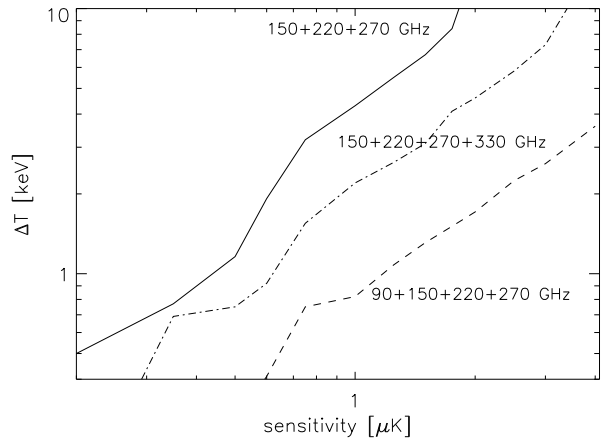}
\end{center}
\caption{$1\sigma$ error-bars on $T_e$ as a function of the observing
sensitivity for a cluster with $T_e=5$ keV, $v_p=200$ km/sec and 
$y=2 \times 10^{-4}$. The solid line is for an experiment with
3 observing frequencies centred at 150, 220 and 270 GHz (ACT-like). 
The dashed and dot-dashed lines include one additional observing frequency,
at 90 GHz or 330 GHz respectively.}
\label{fig:sensigt}
\end{figure}

Our previous results tell us that extracting the electron temperatures
from the data of the studied SZ experiments is problematic due to the
limited sensitivities. We have calculated the error-bars on $T_e$ for
a specific cluster ($T_e=5$ keV, $v_p=200$ km/sec and $y=2 \times
10^{-4}$) as a function of the sensitivity. The result is shown in
\fref{fig:sensigt}. The solid line is for an ACT-like experiment
with 3 observing frequencies placed at 150, 220 and 270 GHz. As one
sees, to reach $\Delta T_e$ of the order 1 keV one must have a
sensitivity of about 0.5 $\mu$K. However, by introducing one
additional channel the errors can be significantly reduced. In
particular with an extra measurement at 90 GHz, one can reach $\Delta
T_e$ of 1 keV (4 keV) already with $1~\mu$K ($4~\mu$K) sensitivity.
We thus emphasize that future SZ observations with several observing
frequencies should be able to extract the electron temperature
directly without the need for independent X-ray information.

Good temperature sensitivities are expected for the next generation of
SZ experiments. For instance the South Pole Telescope (SPT) is a new
and ambitious project~\cite{spt}, that will use the SZ effect to
search for and count 10,000 clusters of galaxies and cover 4000
squared degrees per year. This telescope will consist of a single 8
meter primary dish, accompanied by an array of 1,000 bolometers, and
will measure temperature differences with an accuracy of $0.1 ~\mu$K.
With several observing frequencies and such impressive sensitivity one
will be able to deduce the electron temperature purely from SZ observations.

We remark that SZ can probe the outermost regions of clusters better
than X-ray observations, simply because the SZ effect decreases like
$n_e$, whereas X-ray brightness decreases as $n_e^2$. Thus with direct
SZ temperature observations and good angular resolution one can find
the density profile of the outer cluster region, which directly allows
one to infer the outer dark matter density profile~\cite{hansenstadel}.
This dark matter profile has not been measured yet at such very large
radii, but is predicted from N-body simulations to be $\rho \sim r^{-3}$.

\section{Optimal frequency choice for a SZ experiment}
\label{optimal}

\begin{figure}
\begin{center}
\epsfxsize=9cm
\epsffile{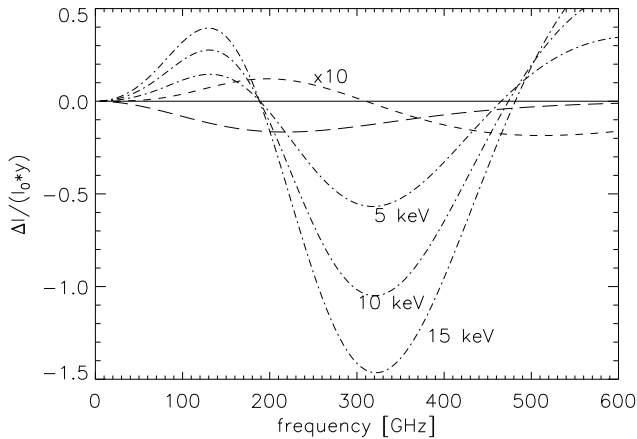}
\end{center}
\caption{The 3 dash-dotted lines are the relativistic contributions to
the thermal effect for temperatures 5, 10 and 15 keV. The short-dashed
line is ten times the relativistic correction to the kinetic SZ effect
(long-dashed line) for $v_p=200$ km/sec and $T_e=10$ keV.}
\label{fig:corrections}
\end{figure}

The question of an optimal choice of observing frequencies for the SZ
effect has been posed a long time ago and is motivated by the fact
that one wants to measure both the thermal and kinetic SZ effects.
The measurement of both effects has great importance for cosmology
since it allows to probe the clusters physics (through thermal SZ) and
the matter distribution at large scales (through kinetic SZ, see for
instance Dor\'e et al.\ \cite{dore02}). In order to attain this
goal, one needs to have observations at several frequencies in the
positive and negative part of the intensity change.  In particular, one
should have an observation near 218 GHz, where the thermal effect
vanishes (\fref{fig:example}, solid line). At this frequency one
can theoretically measure the kinetic effect directly
(\fref{fig:corrections}, dashed line) which happens to be maximal
at this frequency. In a recent paper \cite{hol02} it was pointed out,
that 217 GHz may not be an optimal frequency when one observes with
only 3 frequencies.  We will discuss observations with more than 3
frequencies for which such arguments may not apply.

A future, optimal and dedicated SZ experiment could possibly have many
observing frequencies.  Let us here discuss, from a theoretical point
of view, which frequencies would be preferable. Basically we want to
extract all 3 SZ parameters, and the actual frequency choice depends
upon which parameters we want to extract most precisely, and whether
we have prior information such as a temperature determination from
X-rays.  In order to address the question of optimal frequency choice,
we start by re-computing the frequency dependence of both thermal and
kinetic SZ effects in view of the most recent developments in the
subject. Namely, we compute the thermal effect in its exact form, i.e.
including the corrections for the relativistic effects according to
Dolgov et al.\ \cite{dol01}.  The 3 dash-dotted lines in
\fref{fig:corrections} are the relativistic contributions to the
thermal effect, for cluster temperatures $T_e=5$, 10 and 15 keV.  That
is, for a given temperature one can calculate $\Delta I(T_e)$, and the
3 dot-dashed curves are thus $\Delta I(T_e)-\Delta I(T_e=0)$,
normalized to $I_0y_c$.  As is clear from the figure, this quantity is
approximately zero around 181 GHz and again near 475 GHz. This feature
is almost independent of the temperature in the first case, but not
for the frequency region around 475 GHz.

We have also included in \fref{fig:corrections} the kinetic SZ
effect (long-dashed line) for a peculiar velocity of 200 km/sec, while
the short-dashed line is the small relativistic correction to it
\cite{noz98,saz98}, enlarged 10 times.  That is, for a given
temperature and peculiar velocity one can calculate $\Delta I_{\rm
kin} (v_p, T_e)$, and the short-dashed curve is thus, $\Delta I_{\rm
kin} (v_p, T_e) - \Delta I_{\rm kin} (v_p, T_e=0) $. The particular
case shown was calculated for $T_e=10$ keV. Thus, if one wants to
differentiate between variations in the background CMBR signal and the
kinetic SZ signal, then one should use these corrections, and place
observing frequencies at the zero and extrema of the short-dashed
curve. The relativistic correction to the kinetic effect is
proportional to $v_p T_e$, and for a 10 keV cluster with very large
peculiar velocity 1000 km/sec, this effect is about a factor 10
smaller than the relativistic correction to the thermal effect. We
have seen that the temperature roughly can be detected through the
relativistic correction to the thermal effect with an observing
sensitivity about 1 $\mu$K, so one should have at least 0.1 $\mu$K
sensitivity (the expected sensitivity of SPT) in order to distinguish
between variations in the background CMBR signal and the kinetic SZ
effect.


First of all, to extract the dominating comptonization parameter
$y_c$, one should pick a frequency where the relativistic corrections
to the thermal and the kinetic SZ effects give small contributions.
Clearly, which is well known, the smaller frequencies are good in this
respect (see e.g.\ \cite{hol02} for a recent discussion).
Thus if one wants to remain in the atmospheric window, then a
frequency of 90 GHz is good (a smaller frequency is even better if
the point sources can be acceptably removed). It turns out, however,
that an equally good frequency is near 475 GHz
(\fref{fig:corrections}). At this frequency, the contributions from
both the kinetic SZ effect and the relativistic corrections are very
small, while at the same time the thermal SZ effect is still quite
significant. Observations at such large frequencies would require a
careful removal of dust contamination, which can be achieved by fitting
the observations at several even higher frequencies to a dust model.
The resulting fit can be extended down to lower frequencies in order
to extract the dust from the signal \cite{pgb98,lam98}.

Let us now concentrate on the cluster temperature. One can in theory
extract the temperature using the relativistic corrections to the
thermal effect. To do so one should choose frequencies where the
relativistic contribution (dot-dashed lines in \fref{fig:corrections})
is maximal and zero~\cite{melch2002}.  The maxima are near 120 and 330
GHz, and as discussed above the zero points are at 181 GHz and near
475 GHz. On the other hand, in order to extract the peculiar velocity
one has to measure the kinetic SZ effect (dashed line,
\fref{fig:corrections}) which reaches its maximum near 217 GHz. The
relativistic corrections to the kinetic SZ effect (long dashed line,
\fref{fig:corrections}) are maximal near 200 and 510 GHz, and
disappear near 310 GHz.

\begin{figure}
\begin{center}
\epsfxsize=9cm
\epsffile{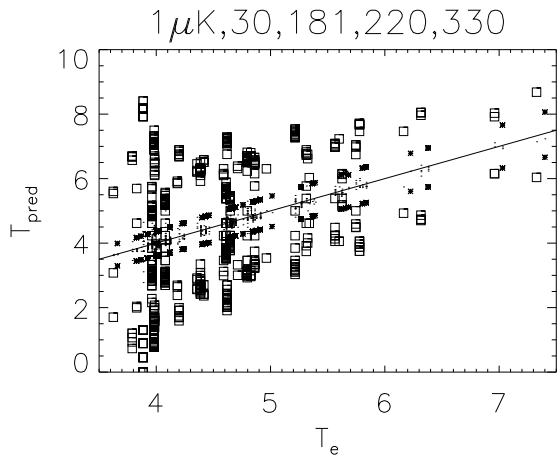}
\end{center}
\caption{A hypothetical experiment with sensitivity of 1 $\mu$K, with
only 4 frequencies placed for optimal temperature determination. Open squares
are with no prior temperature knowledge, stars are with $10\%$ temperature
information.}  
\label{fig:t.30.181.220.330}
\end{figure}

As a specific example of the parameter extraction, let us consider an
experiment with 4 frequencies and sensitivity of $1~\mu$K. Let us try
to extract the temperature as well as possible. Due to the importance
of the comptonization parameter we choose 30 GHz (assuming that point
sources can be removed). For the temperature itself we choose the
maximum and zero of the dot-dashed lines, namely 181 and 330 GHz. To
remove the effect of peculiar velocity we include as the last
frequency 220 GHz.  The results are presented in
\fref{fig:t.30.181.220.330}. One finds that for a cluster of 5 keV and
comptonization parameter of $3 \times 10^{-5}$, that with only 4
frequencies, one can extract all 3 parameters simultaneously with
$1\sigma$ error-bars (open squares) of about 2 keV, $10^{-6}$ for
$y_c$, and $v_p$ about $50\%$. For larger cluster temperatures the
temperature will be extracted with better precision, since the
relativistic corrections to the thermal SZ effect are more important.
With prior knowledge about the temperature of $10\%$ (stars), then the
precision on $y_c$ is improved by almost a factor of 2, and the
precision on $v_p$ with almost a factor of 5. It is therefore clear
again, that prior temperature knowledge is very important and should
be included if one considers small cluster samples.  It is even
crucial when one wants to deal with the proper motion of the galaxy
clusters and the large scale motions.

\section{Bandwidth and precision of measurements}
\label{bandwidth}

We have so far been discussing how well a given experiment with few
$\mu$K sensitivity can extract the cluster parameters, however, for
this discussion to be relevant we must make sure that $\mu$K sensitivity
is achievable.  As we will see below, systematic errors as large as
$\mu$K can appear simply through the finite bandwidths, and such
systematic errors cannot easily be removed even in principle.

The distortion of the CMBR caused by the SZ effect can be formally
defined through an effective change of temperature for each frequency
\begin{equation}
\Delta T (x) = T_0 \frac{\Delta I (x)}{I_0} \frac{(e^x-1)^2}{x^4 e^x}~,
\label{dT_DI}
\end{equation}
which is only a formal representation of the SZ distortion, since
strictly speaking temperature is not a well defined quantity when the
photons are not in equilibrium.  The above expression is derived from
the relation $\Delta I (x)=I_0 x^3 (f-f_0)$ assuming that $x \Delta T
(x)/T_0 \ll 1$, and can have corrections of the order $x \Delta T
(x)/T_0$, which can be important for precise measurements. However, we
will consider eq. \eref{dT_DI} as an exact definition of the
effective temperature difference in order to avoid confusions.
\begin{figure}
\begin{center}
\epsfxsize=9cm
\epsffile{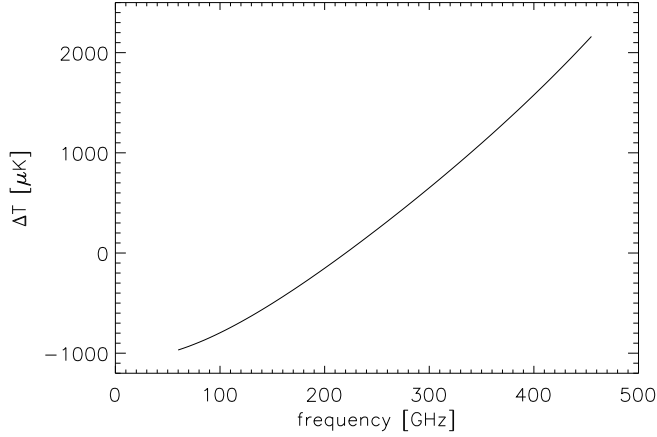}
\end{center}
\caption{The SZ effect in terms of an effective $\Delta T $ from
  \eref{dT_DI} for $y_c=2\times 10^{-4}$, $T_e =5$ keV and
  $v_p=-200$ km/sec. }
\label{fig:dT_total}
\end{figure}

\begin{figure}
\begin{center}
\epsfxsize=9cm
\epsffile{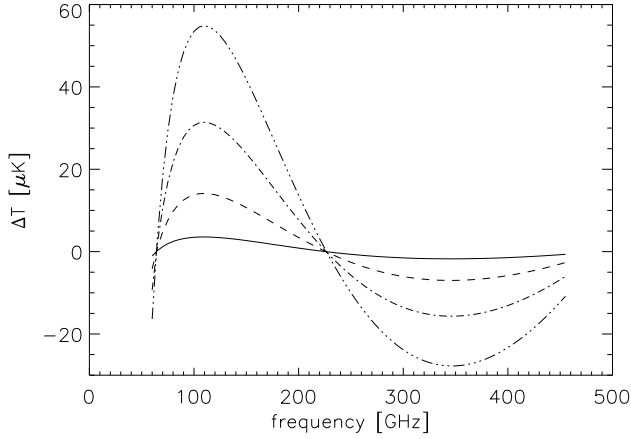}
\end{center}
\caption{Shift in the non-relativistic SZ effect from finite bandwidth
  ($\delta \nu = 10 , 20 , 30$ and $40$ GHz) with a broad top-hat
  filter, if the shape of the SZ effect is ignored.  The 40 GHz filter
  gives the larger effect.  Parameters are the same as in
  \fref{fig:dT_total}.}
\label{fig:dT_f0}
\end{figure}

The SZ effect in terms of temperature difference from eq. 
\eref{dT_DI} is shown in \fref{fig:dT_total} for cluster
parameters $y_c=2\times 10^{-4}$, $T=5$ keV and $v_p=-200$
km/sec.  The typical values of $\Delta T \sim 100 - 500 ~\mu$K are
more than two orders of magnitude larger than the desired precision of
future experiments, which are of the order $0.1-5~\mu$K. In this section, we
investigate the effects of a finite frequency bandwidth, which can be
comparable to, or even larger than, the expected error-bars of future SZ
experiments.

Let us start with the SZ effect in the non-relativistic approximation,
and emphasize the well known fact that one must use the exact shape of
the SZ effect when reducing the observations in a given bandwidth to
just one frequency.  Usually the experiments detect all photons which
fall inside of a bandpass with finite width $\pm \delta \nu$ around a
central frequency $\nu_{\rm mid}$ and extrema $\nu_{\rm max}$ and
$\nu_{\rm min}$. However, because the energy of photons in the
atmospheric window is unknown, the resulting flux should be rescaled
to the central frequency. In the non-relativistic case this is simple,
because the functional shape of the SZ effect is known,
\begin{equation}
\frac{\Delta I_{\rm exp} (\nu_{\rm mid})}{I_0} = y_c 
\left(   f_0(\nu_{\rm mid}) +  \epsilon_0 (\delta \nu)  \right)~,
\label{band_dI_NR}
\end{equation}
where $f_0$ is given by \eref{f0} and $\epsilon_0 (\delta \nu) $ is
the  shift due to a non-zero bandwidth
\begin{equation}
\epsilon_0 (\delta \nu) = \frac{1}{\nu_{\rm max}-\nu_{\rm min}}
\int_{\nu_{\rm min}}^{\nu_{\rm max}} d\nu~ 
\left(f_0(\nu)  - f_0(\nu_{\rm mid})\right) ~.
\label{band_eps0}
\end{equation}

There are two possible sources of error in
\eref{band_eps0}. First, the central frequency might not be at
the exact centre of the band, $\nu_{\rm mid} \neq (\nu_{\rm
min}+\nu_{\rm max})/2 $. It was recently shown that bandpass errors at
the 10\% level are acceptably small for Planck \cite{chu02}. Second,
the band can be wide enough that the shape of the SZ effect becomes
important.  We show in \fref{fig:dT_f0} what the error would be if
one does not take the shape of $f(x)$ into account.  The curves
correspond to $\delta \nu = 10 , 20 , 30$ and $40$ GHz.  Fortunately,
such error can be taken into account exactly from equations \ref{f0},
\ref{band_dI_NR} and \ref{band_eps0}.

\begin{figure}
\begin{center}
\epsfxsize=9cm
\epsffile{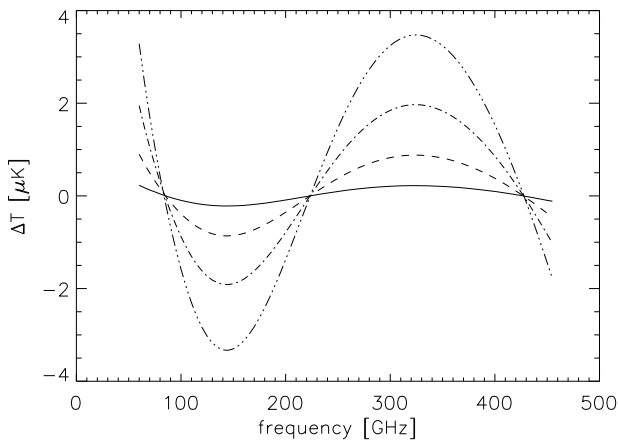}
\end{center}
\caption{The systematic error in the SZ effect from finite bandwidth
  with a broad top-hat filter that arises from the shape of the SZ
  effect when taking into account the relativistic corrections for
  $y_c=2\times 10^{-4}$ and $T_e=5$ keV ($\delta \nu = 10 , 20 , 30$
  and $40$ GHz top-hat filter, where the 40 GHz filter gives the
  larger effect).}
\label{fig:dT_df_T5}
\end{figure}

Now, let us consider the case including relativistic corrections,
where the exact shape is a priori unknown,
\begin{eqnarray}
\frac{\Delta I_{\rm exp} (\nu_{\rm mid})}{I_0}  
&=& y_c \left( \right.f_0(\nu_{\rm mid}) + \epsilon_0 (\delta \nu) 
\nonumber \\ &&+ \left.
\delta f(\nu_{\rm mid},T_e) + \epsilon_1 (\delta \nu)\right)~,
\label{band_dI_tot}
\end{eqnarray}
where $\delta f(\nu,T_e)$ are temperature corrections and
\begin{equation}
\epsilon_1 (\delta \nu) =  \frac{1}{\nu_{\rm max}-\nu_{\rm min}}
\int_{\nu_{\rm min}}^{\nu_{\rm max}} 
d\nu ~\left(\delta f(\nu,T_e)  - \delta f(\nu_{\rm mid},T_e)\right)
\label{band_eps}
\end{equation}
is the bandwidth error from the relativistic corrections to the
thermal SZ effect.  This error is shown for bandwidths $\delta \nu =
10 , 20 , 30$ and $40$ GHz in \fref{fig:dT_df_T5}. From this
figure we conclude that, if the experimental precision is better than
$\sim 5~\mu$K and the frequency bands are large enough, one should take
into account the correction from \eref{band_eps}. This error will
increase for larger temperatures and may also have important
contribution from non-zero peculiar velocities, which will shift the
shape and zero-point of \fref{fig:dT_df_T5} slightly.  The main
problem of taking into account the error $\epsilon_1 (\delta \nu)$ in
\eref{band_eps} is that the electron temperature is a priori
unknown.  Thus, by using the non relativistic form in eq. \eref{f0}, to
reduce the observations in a given bandwidth to one single frequency,
one will always have a systematic error as large as shown in
\fref{fig:dT_df_T5}. The simplest way to avoid such error is to
reduce the bandwidth, but this would lead to a reduction in the signal
to noise and would increase the statistical error-bars. A possible
procedure to significantly reduce this systematic errors could be the
following. First, reduce to one frequency, taking into account only
$\epsilon_0$ from \eref{band_eps0} (that is, assume the
non-relativistic form).  Afterwards, one finds the best fit point in
the parameter space ($y_c$,$v_p$,$T_e$). Finally, one can recalculate
everything, taking into account $\epsilon_1 (\delta \nu)$ for the
reduction to one frequency (that is, using the real form of the SZ
effect including temperature corrections and peculiar velocity), and
subsequently find the new best fit point values for the SZ parameters.

\section{Conclusions}
We have pointed out that there are parameter degeneracies for the
Sunyaev-Zel'dovich effect. This in particular implies that the
error-bars on the peculiar velocities for clusters with positive
peculiar velocities is significantly smaller than for cluster with
negative peculiar velocities. This parameter degeneracy can be broken
either by having prior temperature knowledge, or by observing at
several frequencies. 
For an experiment with $2~\mu$K sensitivity and 3 observing frequencies
(ACT-like), we find that a Compton parameter of the order of $10^{-4}$
is often deduced with a good accuracy 2\% (10\%) with (without) prior
information on the electron temperature. The prior information on the
electron temperature is even more important for the extraction of the
peculiar velocity for which the accuracy reaches 10--20\%. Additional
observing frequencies also break the parameter degeneracies but the
accuracies are not better than with the prior temperature knowledge.
In the particular case of Planck, the Compton parameter of $~10^{-4}$
is extracted very accurately (2\%) even without information on the
temperature. Prior temperature knowledge is, however, necessary to go
from a peculiar velocity determination of a factor of a few to
20--50\%.

We have discussed the needed sensitivity in order to extract the
cluster temperature using SZ observations only. We have seen that
experiments with only 3 observing frequencies need very good
sensitivity ($\sim 0.5 ~\mu$K) in order to determine the cluster
temperature with 1 keV error-bars. Already with 4 frequencies one can
reach 1 keV (4 keV) temperature error-bars with only $1~\mu$K
($4~\mu$K) sensitivity.

We have identified the frequencies which are optimal for deducing the
3 cluster parameters from a theoretical point of view.  This optimal
choice of 4 observing frequencies and sensitivity of 1 $\mu$K allows
one to deduce the parameters of a 5 keV, $y_c=3\times 10^{-5}$ cluster
with $1\sigma$ error bars of about 2 keV and $10^{-6}$. The velocity
is extracted with a 50\% accuracy.  The optimal frequencies for an
actual experiment will depend both on the sensitivity of the
experiment and also on which cluster parameters are of primary
interest.

We have studied the systematic error which appears due to the finite
band-width.  This error is related to the shape of the SZ effect. The
exact shape of the SZ effect depends on the cluster temperature and
peculiar velocity, which cannot be known a priori. This systematic
error can be of the order of a few $\mu$K, and must therefore be
considered seriously by future experiments which aim at such
impressive sensitivity.

\ack It is a pleasure to thank Lloyd Knox, Francesco Melchiorri,
Fran\c cois Pajot and Rocco Piffaretti for useful discussions. The
work of DS was partly supported by the Deut\-sche
For\-schungs\-ge\-mein\-schaft under grant No.\ SFB 375. SP was
supported by the Spanish grant BFM2002-00345 and a Marie Curie
fellowship under contract HPMFCT-2002-01831.

\label{lastpage}

\end{document}